\documentstyle[11pt]{article}                                               
\begin{document}
\tableofcontents
\pagebreak
\Large
\noindent {\bf What kind of science is cosmology?}
\normalsize

\vspace*{1cm}

\noindent Hubert F. M. Goenner\\

\noindent Institut f\"ur Theoretische Physik - Universit\"at G\"ottingen\\
Friedrich-Hund-Platz 1\\
37077 G\"ottingen
 
\vspace*{2.5cm}

 \Large {\bf Abstract}\normalsize\\

In recent years, by theory and observation cosmology has advanced
substantially. Parameters of the concordance or $\Lambda$CDM
cosmological model are given with unprecedented precision (``precision
cosmology''). On the other hand, 95\% of the matter content of the
universe are of an unknown nature. This awkward situation motivates
the present attempt to find cosmology's place among the (exact) natural
sciences. Due to its epistemic and methodical particularities, e.g.,
as a mathematized historical science, cosmology occupies a very
special place. After going through some of the highlights of
cosmological modeling, the conclusion is reached that knowledge
provided by cosmological modeling cannot be as explicative and secure
as knowledge gained by laboratory physics.

\vspace{1cm}


\pagebreak

\section{Introduction}

\vspace*{1.5cm}

In the past two decades, cosmology has taken a promising course. Due to improved
and new observational instruments and the observations made with them, a
wealth of data has made possible the determination of cosmological parameters 
with higher precision than ever before (``precision cosmology''). On the 
theoretical side, the interaction of elementary particle physicists and 
astrophysicists has provided major contributions to the interpretation of
observations. In spite of the progress made, the standard cosmological model, 
developed into the ``concordance model'', seems not to be in good shape. With 
95\% of the matter content of the universe presently being of an unknown 
nature, can any claim be made that today's cosmological model leads to a better 
understanding of the universe than the model of two decades ago? 

In this situations it may not be contraproductive to inquire about the
nature of the discipline. Here, we encounter a common endeavour of mathematics, 
theoretical physics, astronomy, astro-, nuclear and elementary particle
physics with the aim of explaining more than the cosmogonic myths of our
forefathers. Has cosmology become a natural science, even a branch of the
exact sciences? It certainly is a field of research well established by all 
social criteria if we follow J. Ziman \cite{Ziman1968} and define natural science 
as an {\it empirical} science steered by public agreement among scientists. In 
this context, ``empirical'' means that conclusions are not merely drawn by 
rational thinking as in the humanities but that they are tested by help of 
reproducible {\em quantitative} experiments/observations. Data from these 
measurements are interpreted by consistent physical theories and receive a 
preliminary validation to be reconsidered in the light of new facts. 
Cosmology as a very young scientific discipline has not yet reached
the same degree of differentiation as other subfields of physics.\footnote{This is reflected by the PACS-classification which
    provides only 7 subclasses for cosmology, 20 for solar physics,
    and 178 for ``solid earth physics''.} Most of what follows will refer
to {\em physical} cosmology on a solid empirical basis and to its
subfield named here {\em originative} cosmology. In the latter, the 
speculative parts, {\bf necessarily} implied by physical theorizing,
are dominant; they are just beginning to be linked to empirical
testing or still await probing in the future.\footnote{ An
  endeavour purporting to belong to physics but without any connection
  to an empirical background will be called {\em make-believe}
  cosmology, cf. section \ref{subsection:multiverse}. This is to
  function as a reminder that the universe exists not just ``on paper'' as
the philosopher P. Val\'ery would have it.} 

Three periods of extremely unequal duration in the time evolution of
the expanding universe will be used for gaining an impression of
cosmology. They are: The flashlike ``very early universe'' of $\Delta
t \sim 10^{-12} s$ duration (before the assumed electroweak phase
transition); it includes the inflationary era and {\bf prior}
Planck-scale modeling (quantum cosmology). Next, the ``early
universe'' (until early structure formation) amounting to $\sim 4\%$ of the 
total age of the universe [$(13.27\pm 0.12)\cdot 10^{9}~y$] and covering 
$\Delta t \sim 4.3\cdot 10^{8}$ years; here, nucleosynthesis and the
release of cosmic background radiation (CMB) can be found. Finally,
the remaining period from structure formation (reionization) until
today comprising $\sim 96\%$ of the time. Einstein's theory of
gravitation will be the almost exclusive theoretical background 
adopted here because its implications for physical cosmology have been developed
best. In the following, I shall use the words ``cosmos" and ``universe'' as 
synonyms although they carry different rings; cosmos goes well with order and
coherence, while universe implies uniqueness and entirety. Before going into 
details of cosmological modeling I will try to circumscribe cosmology as a 
field of research.

\section{The content of cosmology}
\label{section:content}
\subsection{The universe: a well defined physical system?}
Sciences or branches of science are classified by the subject
investigated, or by the methods of investigation used. Thus, cosmology
could be called ``cosmophysics'' in parallel with geophysics or solid
state physics because its subject is the cosmos. In this spirit, in
dictionaries, cosmology is defined as the general science of the universe
\cite{FunkWagnall1974}, the science of the physical laws of the universe
\cite{PetitRobert1985} or, as the Oxford Companion has it: ``the study of the 
entire Universe'' (\cite{LiddleLoveday2008}, p. 61). A textbook tells
us: ``In cosmology we try to investigate {\it the world as a whole}
and not to restrict our interest to closed subsystems (laboratory, Earth, 
solar system etc.)'' \cite{SexlUrbantke1983}. The world {\it as a whole}, 
though, is not readily accessible, empirically. Whether bootstrap definitions 
like the universe is  ``the largest set of objects (events) to which physical 
laws can be applied consistently and successfully''
\cite{Bondi1961}, or formulations as ``the universe means all that 
exists in a physical sense'' (\cite{Ellis2006}, p. 1) are more helpful, is a 
matter of taste. Once in a while, even a religious flavour is added when the 
universe is ``usually taken to mean the totatility of creation.'' 
(\cite{Carr2007}, p. XV). 

In this situation, scientists provide qualifying attributes, and point to
subfields of cosmology linked with them \cite{Ellis2006}): the {\it observable}
universe, the {\it visible} universe, the {\it physical} universe, the {\it 
astronomical} universe \cite{McVittie1961}, the {\em astrophysical} 
universe \cite{Peebles1993}. Although, at present the {\em
  biosphere} is  not included in cosmology, by some of these
attributes it is not strictly ruled out. In order
to be able to do physics, an idealized subsystem of ``all that exists'' must
be selected. A preliminary definition, i.e., ``we understand the universe to
be the largest presently observable gravitationally interacting system'', would 
satisfy the needs of the practizing cosmologist.\footnote{Gravitation is the 
dominant interaction on the largest scales. On smaller scales, all other 
interactions come into play.} From the point of view of epistemology, such a 
definition is hardly acceptable, though. The observable universe changes 
permanently, because the domain of nature observable to us depends on the
power of the available measuring instruments. Consequently, a further
definition of the ``observable universe'' reads as ``what in principle we can 
observe'' (\cite{LiddleLoveday2008}, p. 314). Cautious authors have avoided the 
word ``universe'' altogether in favor of expressions like ``the metagalaxy'' 
\cite{Alfven1967}, ``distribution of matter on the largest scale'' 
\cite{Buchdahl1981}, or ``structure on a large scale'' (cf. \cite{Ellis2006}). 

In spite of this situation, most cosmologists seem not to worry about the 
domain of application of their theories: in the wake of time they expect to
find out. They take it for granted that the physical system ``universe'' is 
as meaningful as the {\em alterable} and {\em touchable} physical 
systems investigated in the laboratory.\footnote{Untouchable physical 
systems as the Sun, a star, a galaxy exert direct sensorial reactions on us. 
The universe does not.} Possibly, the cosmos is definable only in the 
sense of a 
mathematical limit process. Or, as an ontological construct: ``the largest 
inextendible  entity''. Progress of research seems not to be hampered by this
attitude.\footnote{In this spirit, in recent monographs the physical system 
``universe'' remains undefined (cf. \cite{Mukhanov2005}).} In comparison, the
concept of elementary particle is accepted in the sense of 
the smallest indivisible entity. At first, it should have been the atom, then 
the nucleus and, presently, it is the quark - with no end of further 
subdivisions in sight. An approximative definition of the universe as
a physical system may well be the only one allowed to physicists; however, there 
is the danger that the epistemological background gets out of sight. In fact, 
particularly in quantum cosmology and in approaches related to string theory, 
the universe is treated as an entity resembling more a particle among other 
particles than the totality of gravitationally interacting masses on the
largest scale (Cf. also sections \ref{subsubsection:peculiar} and
\ref{subsection:multiverse}). In a way, methodologically, cosmophysics is {\it
  opposite} to phenomenological thermodynamics. There, valid laws are
formulated without the need of knowing the detailed microscopical structure of 
matter. In cosmophysics, until recently (dark energy!) we were dealing
with the detailed knowledge of structured parts of a system unknown in
its totality. 

If cosmology were just a branch of {\it applied mathematics} we could
{\it define} it as the study of the global properties of ``cosmological 
solutions'' of certain field equations, notably Einstein's
(cf. \cite{HawkingEllis1973}). We would then include singularities
(e.g., at the big bang) as 
boundary points of the Riemannian manifold representing the universe. However, 
the qualification of an exact solution as a model for the cosmos still would 
have to be made by borrowing ideas from physics; for example, by the kind of 
isometry group to be assumed. Possibly then, homogeneous and isotropic
cosmological models with {\it compact} space sections of {\it negative} 
curvature would have to be discarded because they admit only a 3-parameter 
isometry group, globally \cite{Ellis1971}.\footnote{Cf. also, cosmological
  models with multiply connected space sections (\cite{EllisSchreiber1986}, 
\cite{LachiezeLuminet1995}, \cite{Luminetal2003}, \cite{Aurichal2008}} The 
cosmological models of applied mathematics which, by careless use of language 
sometimes were called ``cosmologies'' (\cite{Ellis1991},
\cite{Halliwell1991}, \cite{RyanShepley1975}) or ``universes''
(\cite{Leeuwenal1989}, \cite{RobertsonNoonan1968}), need not have any relation 
to the world outside of our brains. This point is not a side issue: in the
``multiverse scenario'' no distinction is made between what is a mental 
construct and what, by its relation to empirical data, can be accepted as some 
kind of ``reality'' external to our mind, cf. section
\ref{subsection:multiverse}.  

\subsubsection{A mathematized historical science?}
\label{subsubsection:peculiar}

With astronomy, cosmological research shares the situation that its object,
the universe, or parts of it of cosmic relevance, have to be observed
at a space-time distance, measured on and inside the {\em past} lightcone 
from a tiny part of the Earth's (or the solar-system's) worldline. Experiments 
cannot be carried out for observing effects. Observational cosmology may be 
compared to geological, palaeontological or archeological field work: deeper 
and deeper strata of the past are excavated, with the difference to 
palaeontology and archeology being that the present state of the objects 
observed is unknown. Cosmological theory does not describe a museum of relics 
but a dynamical system. Also, for cosmology, better mathematical models
  exist. 

This historical aspect is not the full story, but it shows up in many ways; 
one of them being the transformation of the concept ``prediction''. In 
cosmology, without exception, prediction is a conclusion from present 
observations to {\it past} times or, vice versa, after hypothetical input for 
past times, to consequences for the {\it present}. In slightly altering a 
statement of Friedrich Schlegel (who directed it toward historians):
cosmologists are prophets for the past. In physics proper, prediction
means the foretelling of a {\it future} state from conditions given
{\it now}. The social usefulness of natural science (and technology)
rests on this regular meaning of prediction. Certainly, cosmological
models can be used to make exact {\em calculations} toward the future 
\cite{Dyson1979}, \cite{Discussal1985}. For cosmological time
  scales these calculations are pointless, however, because they
cannot be validated by observational tests: Will any of them be preserved 
for a test in $\sim 10^6$ years? Even if cosmological theory could
provide us with a reliable description of the past, its validity for the 
future cannot be probed; it is a consequence of continuity assumptions for 
the mathematical equations of theoretical cosmology. If the precision
of, say, spectroscopic measurements could be increased to the extent
that the changes in redshift of distant objects can be monitored over
a time-span within our lifetime, then extrapolations applicable to the 
motion of the objects, for the {\em near} future only, will become
possible. In ``make-believe cosmology'', the ``ultimate fate of the
Universe'' is broadly discussed with future events timed with little 
reservation (cf. \ref{subsection:multiverse}).

A sober physical and philosophical assessment of a ``lack of predictability in the
real universe'' is given by (\cite{Ellis2007a}, p. 61).  

Nowadays, the word ``prediction'' is used by most physicists working in
cosmology as meaning ``a consequence of'' without any implication of linking 
the present to the future. This can become rather quixotic as in: ``[..], a 
fundamental discreteness of spacetime at the Planck scale of $10^{-33}$ cm 
seems to be a prediction of the theory [..].'' 

\subsubsection{Other features peculiar to cosmology}
\label{subsubsection:peculiar2}

A characteristic feature of the universe, once believed to be important, is
its {\it uniqueness}: one and only one such physical system (``the world as a
whole'') can be thought of as given to us. Unfortunately, with the
advent of quantum cosmology and superstring theory, a semantical erosion of 
the word ``universe'' has begun. Already two decades ago, we had been asked 
``How many universes are there?'', when authors investigated ``a dilute gas 
of universes'' or a ``single parent universe ... in a plasma of baby
universes'' \cite{Strominger1991}. We were 
approached to ``suppose universes are emitted from $t = 0$ like photons from
an antenna'' \cite{Susskind1991}. At the time, it remained a miracle, though, 
what kind of tangible receptacle could house or receive multiple universes. 
By now, this problem seemingly has been fixed by the introduction of the 
concept ``multiverse'' (Cf. section \ref{subsection:multiverse}).  

If the uniqueness of the universe is accepted, why then is this system so
special? Isn't the Earth unique, too? True, as far as its {\it
individuality} is concerned. But the Earth is just one of the planets in the 
solar system and one of billions more conjectured around other stars 
(exoplanets).\footnote{The search for exoplanets with parameters
    close to those of the Earth may form a link to the biosphere.} It gets its 
individuality {\it by comparison} with other
planets. In contradistinction, is there an empirical or a conceptual way of 
comparing ``our'' universe to ``others''?\footnote{Of course, {\it
    cosmological models} can be compared with each other - on paper, though.}
In speculations of past years, statistical methods were applied to a set of 
``universes'' residing in the mind in order to get a handle on the
values of fundamental constants of nature \cite{Linde1990}. 

As a consequence of the uniqueness of the universe, {\it specific cosmic laws 
cannot obtain} \cite{Munitz1963}. It is not excluded that new physical laws will 
be discovered while we try to scientifically describe the cosmos. Such laws, 
however, will refer to properties of parts (subsystems) of the universe and 
to relations among them.

Can theories applying to a single object be falsified? The example of
the {\it steady-state} cosmological model seems to show that falsification is 
possible for statements of cosmological theory, because observations made now 
are observations of past states of the universe. Yet, as the complex attempt 
at a revival of the steady-state model shows \cite{NarlikarBurbidge2008}, some 
caution is in order. This, again, indicates that cosmology could be
interpreted as kind of a mathematized {\it historical} science: with 
falsification meaning nothing more than that our interpretation of the 
historical record has been mistaken and must be revised.

\subsubsection{Initial conditions}
\label{subsubsection:initial}

The Einstein-field equations for the cosmological model being {\it
hyperbolic} partial differential equations, a {\it Cauchy initial
value problem} with given initial data must be solved in order that
we may arrive at a unique solution. An additional chain of
argumentation or even a theory must be developed by which the initial
data actually in effect for the universe as we observe it are picked
out from among the imagined set of all possible initial data. Thus {\it
  cosmogony}, the theory of what brought the cosmos into being, and cosmology 
are inseparable \footnote{The assumption of temporal closedness of the 
universe is one escape route in sight. With its painful consequences for 
causality and pre-(retro-) dictability, the
idea has not yet been taken seriously. The idea of a cyclic universe with
multiple beginnings and ends also has been proposed since antiquity. For
recents proponents with very different suggestions,
cf. \cite{Penrose2005}, \cite{Bojowald2008}.}. The rise of quantum
cosmology indicates an attempt of bringing cosmogony into the reach of science 
(Cf. \ref{subsection:quantcos}). 

Already within classical theory, attempts had been made to understand 
homogeneity and isotropy near the big bang \cite{CollinsHawking1973}, 
\cite{Misner1968}. R. Penrose suggested to assume homogeneity of space - 
corresponding to a low value of entropy - as an initial condition. He
tentatively used the Weyl tensor as a measure of the entropy and required it 
to vanish at singularities in the past \cite{Penrose1979}, \cite{Penrose1986}, 
(\cite{Penrose1989a}, p. 344). Moreover, in this context, various {\it anthropic 
principles} (\cite{Carter1974}, \cite{DemaretLambert1994} have been invoked 
since their first formulation, and are used even heavier, today.\footnote{The 
debate is still going on whether anthropic principles are useful as a
selection principle with an exploratory value, or just express a demand for 
self-consistency of the cosmological model.} In fact, within make-believe 
cosmology, the search for a rationale for the initial data required for the 
universe to be as it appears to be, seems to be a main motivation. 

As an aside: a related question is whether observation of the physical 
system ``universe'' will permit, in principle, a reconstruction of its initial 
state. Even for as simple a system as the solar system such a task is rather 
difficult. From what can be learned from deterministic chaos and, in view of the 
possibility that the Einstein field equations need not be an ever-lasting 
foundation of cosmophysics, particularly for what happened right after the big
bang, we should remain reserved in this matter. Fortunately, for the standard
cosmological model, initial data for the very beginning of the universe (at
the big bang) are {\em not} needed. Nevertheless, initial data are required at 
the beginning of the inflationary phase. These may be guessed and
validated in the sense of being consistent with what is derived theoretically 
and then observed (cf. \ref{subsection:inflation}).

The fact that we need initial, not final conditions reflects the open
problem of the {\it arrow of time}: how to derive the unidirection of time 
when the basic equations are time-symmetric? Is it linked to the 
``collapse of the quantum wave function''? (\cite{Ellis2007a}, p. 76; cf. 
however {\cite{Zeh1999}.)

\subsection{Cosmological questionaire}
\label{subsection:question}
\noindent With the beginning of research in cosmology a list of general
questions arose:\\
 - Is {\it space} (defined by the distance range between gravitating bodies)
       of finite or infinite  extension? \footnote{The property of
       being infinite refers to the mathematical model. It has no
       observational meaning.}\\
 - Is {\it time} (defined by the duration of certain systems as compared to 
others) of finite or infinite duration in the future, in the past?\\
 - How does cosmic dynamics look (phases of accelerated and/or decelerated 
expansion, structure formation, etc.)?\\
 - What is the matter content of the universe? In the form of baryons, of 
radiation (zero mass particles), of dark matter? What is dark matter made from?\\
 - Is a non-vanishing cosmological constant needed?\\   
If the system were finite in space and in past time, we might ask for the {\it
  total} mass (energy), angular momentum, electric charge, etc and the
{\it age of the universe}. The last concept is reasonable only if all parts of
the cosmos can be parametrized by one single time parameter. In case there is
a dynamics, the initial state of the universe and its evolution in time are of 
interest. Numerous further questions will arise within the three pieces of 
cosmological modeling to be briefly discussed below. Some believe that, by the 
presently accepted cosmological model ($\Lambda$CDM), many of these questions 
have been brought nearer to an answer (Cf. section \ref{subsection:concord}).

\section{Cosmological modeling}

\subsection{General hypotheses}
\label{subsection:genhypo}
\noindent As far as the universe is traced by its large scale mass
structures (galaxies, clusters of galaxies, superstructures), the
questions asked in observational cosmology are concerned with the angular and 
in-depth distribution of such structures, their material content, the
occurence of chemical elements, the origin of particular objects, as e.g., 
quasars, or galactic nuclei, the strength and time-evolution of magnetic and 
radiation fields, etc. In this respect, the highly isotropic microwave 
background (CMB), a Planck-distribution to temperature $\sim 2.7 K$, 
interpreted to be of cosmological significance, is a very important 
characteristic. From the observations, properties will be ascribed to the 
universe serving as entries for cosmological model building.

As main result, a {\it compatibility} with observations of
cosmological significance had been found: the {\it expansion of the
  universe} (redshift), the {\it isotropy of the slices of equal time}
(CMB), and the ``cosmic'' abundance of light chemical
elements. Isotropy does not refer to the position of the earth, the
solar system or the Galaxy but to an imagined rest system defined by CMB itself. 
Nucleosynthesis calculations lead to a value for the average matter (baryon) 
density of the universe {\it consistent} with what is observed, directly,
from luminous masses and, indirectly, through dynamical effects in galaxies
and clusters of galaxies depending also on dark matter. 

Before a {\em quantitative} description of the universe can be attempted, a 
particular cosmological model, i.e., a metric representing the gravitational
potentials, and a description of its material sources, must be given. In order
to reach a unique model, a number of simplifying assumptions usually is made. The 
historical and epistemological background is provided by what often is called the 
{\em Copernican Principle}: ``The Earth does not occupy a prefered position in
the universe.'' Expressed differently, some kind of {\em homogeneity} of space is
demanded. Mathematically, this is expressed by requiring a transitive group 
of quasi-translations (isometries) to act on spacelike hypersurfaces. This
still leaves a sizable number of cosmological models
(cf. \cite{KraSteMaHe1980}, 
particularly secs. 12.3, 12.4, and 15.3). Also, the Copernican principle is {\em
  untestable} as long as we cannot observe the universe, say, from
another galaxy. It can be tested only along our past lightcone by the
counting of sources as a function of redshift. By transforming redshifts
(look-back times) into spatial distances, homogeneity of space then may be
infered. However, the calculation already must involve a cosmological model.

In order to further reduce the number of cosmological models, the
Copernican Principle is replaced by the {\em Cosmological Principle}:
``The universe must be homogeneous and isotropic.'' Isotropy means
that the rotation group acting on spacelike hypersurfaces is also a
symmetry group. This principle leads to a unique class of cosmological
models (FLRW, cf. section \ref{subsection:morestandard}). It likewise
is not testable from our vantage point in the
universe.}\footnote{Homogeneity follows if the universe is
  isotropic around more than one point in a spacelike hypersurface,
  cf. \cite{Ellis1975}. It is surprising that authors think that
  ``homogeneity on large scales {\em is} an extremely strong
  prediction of $\Lambda$CDM'' (\cite{Hoggal2005}, p. 2) whereas this 
homogeneity is built into the $\Lambda$CDM-model as one of its
fundamental assumptions.} 

From the point of view of what is observed (large scale galaxy structure,
cosmological background radiation (CMB)), the Cosmological Principle can lead 
merely to an approximate description of the universe. A large fraction
of cosmologists starts with the Cosmological Principle and accounts for the 
inhomogeneities of the matter distribution and the minuscule anisotropies in
CMB by superimposing them onto the model via perturbation
calculations. Other cosmologists first apply an averaging over space volumes
to the Einstein equations in order to take account of inhomogeneities. Time
derivation and averaging over space do not commute. The procedure is called
backreaction (of the inhomogenities) and  leads to additional terms in the
usual (Friedman-) equations for the homogeneous and isotropic model. For 
applications and a recent review cf. \cite{Buchert2001}),
\cite{Buchert2008}. Still other researchers
directly start from exact inhomogeneous and isotropic solutions of Einstein's
equations (collected e.g., in \cite{Krasinski1997}) and try to fit them to the 
observations. Suggestions also have been
made for using the Cosmological Principle only as an initial condition
for the development of the Universe \cite{HeckSchue1959}, or for
interpreting it in an average sense (``statistical cosmological
principle'' \cite{Schwarz2009a}).

In the following, we list a few assumptions necessarily leading to
a homogeneous and isotropic cosmological model. These assumptions should be 
testable by their consequences. With better data, they could be relaxed as well.

 - $A_1$ The physical laws, in the form in which they are valid 
here and now, are valid A) everywhere, and B) for all times for which the
  cosmological model is expected to be valid.
Otherwise, it would be impossible to uniquely interpret
  observations. In theory, it would make no sense to apply the standard model of
  elementary particles or Einstein's theory of gravitation to the early
  universe. $A_1$ can also express the hope that local and global physics (of
the universe) are {\it not} inextricably interwoven: ``physics on a small 
scale determines physics on  large scale'' \cite{Ohanian1976}. The
opposite view 
that ``the physical laws, as we usually state them, already involve the
universe as a whole'' gets only a minority vote \cite{HoyleNarlikar1974}.\\ 

- $A_2$ The values ascribed to the fundamental constants here 
and now are the same everywhere and at all times.\\
When speaking about fundamental constants, we naively think of quantities like 
$c$ (velocity of light), $h$ (Planck's constant), $k_B$ (Boltzmann constant), 
$e$ (elementary charge), $G$ (gravitational constant), or of dimensionless 
combinations of them. In order that the atomic spectra from distant
  objects can be interpreted, the fine structure constant must be assumed to
  be the same as in the laboratory. For a proper interpretation of
  gravitational lensing, the gravitational constant must be assumed to be the
  same as in the planetary system. Then, by observation, bounds for an
  eventual change in the fundamental constants can be obtained, in principle. 
Of course, it is the underlying theories which define these quantities to be 
constant or time-dependent: In scalar-tensor gravitational theory,
  the gravitational ``constant'' would be time-dependent by definition. 
For cosmological modeling in the framework of general relativity, $A_2$ is to 
apply for epochs since, and perhaps including, the inflationary
phase. In elementary particle physics, for higher energies fundamental 
constants depend on the renormalization scale. This seems not yet to
play a role for the present cosmological model. Nevertheless, effects
of a running cosmological and gravitational constant on the evolution
of the universe were studied in \cite{GranSol2010}.\\
 
- $A_3$ The universe is connected (in the mathematical sense).\\
As we know from the occurence of horizons, $A_3$ cannot be sharpened to the 
demand that communication is possible between any two arbitrarily chosen
events in the universe.\\ 

- $A_4$ In a continuum model, the material substrate of the universe 
   (including dark matter) is described by a mixture of {\it ideal} 
   fluids - not {\em viscous} fluids.\\
- $A_5$ The material substrate of the universe evolves in time as a 
   {\it laminar} flow - not a turbulent one.\\
The assumption of an ideal fluid without shear and rotation of the 
streamlines as expressed by $A_4, A_5$ uniquely leads to the FLWR-class of 
cosmological models. In $A_4$, an ideal fluid is characterized by the
equation of state $p=w~\rho$ with a constant $1 > w > 0$.
\footnote{Here, $p$ is the pressure and $\rho$ the energy density of the ideal
  fluid. Both, the constancy of $w$ and the range of values allowed will be 
relaxed.} In fact, as an addition to the current standard cosmological
model, effects of viscosity and turbulence in the course of the evolution of 
large-scale structures are being investigated (perturbation theory), e.g., in 
connection with dark matter, or dark energy, and magnetic fields of
   cosmological relevance, etc.\\

$A_5$ also expresses the possibility of a slicing of space-time into 
hypersurfaces of constant time. A fundamental hypothesis going into the
standard model is the concept of a cosmic time {\it common} to all parts of
the universe. In some cosmological models as, for example, in G\"odel's, the 
local spaces of simultaneity are not integrable to one and only one 3-space of
``simultaneous being''. (Cf. section \ref{philoprob}.) 

\subsubsection{Primordial Nucleosynthesis}
\label{subsubsection:nucleo}
Primordial nucleosynthesis is considered to form one of the pillars of the
standard cosmological model. Nucleosynthesis for the light elements $d,~
^3He,~^7Li$, except for $^4He$, depends sensitively on a single
parameter of cosmological relevance entering: the ratio 
$\eta = n_B/n_{\gamma}$ of the number of baryons to the number of photons in 
the universe. $n_{\gamma}$ can be calculated from the microwave
background. The decisive nuclear physics parameter is the neutron's lifetime. 
Because the production of $~^4He$ depends on the number of existing 
neutrino families, it is possible to obtain an estimate consistent with what 
has been found with the largest particle acclerators
\cite{Walkeral1991}. Nevertheless, a recent measurement of the $^4$He
abundance ``implies the existence of deviations from standart big
bang nucleosynthesis'' \cite{IzoThu2010}. 

As to the comparison with observations, except for $^4 He$, for
nine reliable determinations of $^3He$ from high redshift quasistellar
sources, and for seven reliable determinations of deuterium at high
redshifts and low metallicity, the {\it observed} distribution of the
light elements comes from measurements within the solar system and the
Galaxy. The uncertainties are in the range of $0.2$\% for $^4 He$, $5-10$\%
for $d,~ ^3 He$ and $15$\% for $^7 Li$ \cite{VilDol2003}, \cite{Steigman2009}. 
There also remains an unexplained difference between the observed and
the theoretically calculated values for the abundance of $^7Li$. From these
data a $5$\% determination of the baryon density is obtained
\cite{Steigman2009}. There are also observations of the chemical
abundance in very old stars \cite{Frebel2008}, but their cosmological 
relevance is not yet clear. In addition to the restricted
observation-volume, the empirical basis for the abundance of chemical
elements thus is less secure than one might wish it to be. The
comparison of calculated and observed abundances depends highly on 
astrophysical theory (models for the chemical evolution of galaxies and stars).

\subsubsection{Empirical situation with regard to $A_2$}
\label{subsubsection:empirical}
All we can safely claim today, with respect to $A_1$ and $A_2$, is that they
are not in conflict with the empirical data. Reliable such data about a
time dependence of the fundamental constants are still lacking, although much
progress has been made. For the quantity looked at most often, i. e., 
$\dot{G}/G$, bounds between $|\dot{G}/G|~\leq ~10^{-10}~ y^{-1}$
and $|\dot{G}/G|~\leq~10^{-13}~ y^{-1}$ have been derived from various 
investigations (solar system, radar
and laser ranging to moon/satellites, astro-seismology, binary pulsar, big
bang nucleosynthesis, Ia supernovae). Cf. the review by Garc\'ia-Berro et al. 
(\cite{GarciaBerro2007}, p. 139-157). Most of the estimates are dependent
on the cosmological
model. Also, they suffer from short observation spans: measurements in the
solar system cover the past 200 - 300 years \cite{Will1981}. At best, the 
observation time could be extended to $\sim 10^9$ y, i.e., the lifetime of the 
solar system. Only then would this be comparable to Hubble time 
$t_0=\frac{1}{H_0} \simeq 9,77\frac{1}{h}\times 10^9 y$, with $H_0= 100~h~ km
s^{-1}(Mpc)^{-1}$, the Hubble constant measuring present expansion.\footnote{The
  Hubble constant is the present value of the Hubble parameter 
$H(t):=\frac{\dot{a}}{a}$ where $a(t)$ is the scale function of the
homogeneous and isotropic universe model. The dot means time derivation.} The situation is not
better for the estimates on $\dot{G}/G$ made from primordial nucleosynthesis
(PN) giving a value for the ratio of $\frac{G_{PN}}{G_0}= 0.91 \pm
0.07$ taken at the time of big bang nucleosynthesis and at
present. For CMB $\frac{G_{CMB}}{G_0}= 0.99 \pm 0.12$, i.e., since 
$\sim 3~ 10^5$ years. \cite{Steigman2009}. 

As to the determination of upper bounds for the fine structure
constant $\alpha$, constraints coming from terrestrial (Oklo natural
reactor), high-redshift quasar absorption systems, big bang
nucleosynthesis, and the angular spectrum of cosmic background
radiation ``do not provide any evidence for a variation of $\alpha$'' 
(cf. \cite{GarciaBerro2007}, p. 139). Typical results are 
$\frac{\Delta \alpha}{\alpha} = (-0.3 \pm 2.0)\times 10^{-15}~
y^{-1}$ (laboratory), $\frac{\Delta \alpha}{\alpha} =
(0.05\pm 0.24)\times 10^{-5}$ (quasars at $z= 1.508)$),
and $\frac{\Delta \alpha}{\alpha} = (-0.054\pm
0.09724)$ (CMB at $z \simeq 10^3$). Another interesting target has 
been the ratio of proton and electron mass $\mu=\frac{m_p}{m_e}$. A typical
 bound is $|\frac{\Delta \mu}{\mu}| = (-5.7\pm 3.8)\times 10^{-5}$ for
  redshifts of $ z=2.377$ and $z=3.0249$, respectively. (cf. 
\cite{GarciaBerro2007}, p. 159).

The time-independence of the fundamental constants which is particularly 
important in the inflationary phase, is not directly testable during this
period. 

\subsubsection{Cosmological observation}
\label{subsubsection:cosmobs}
In addition to fundamental suppositions for {\it theoretical }
modeling, hypotheses for the gaining of data and the empirical
testing of cosmological models are necessary. Such are, for example:\\
 - $B_1$ The volume (spatial, angular) covered by present
observation is a {\it typical} volume of the universe.\\
The application of $B_1$ may become problematic because of the occurence of 
{\it horizons} in many of the cosmological models used. There may be parts
of the universe not yet observable ({\it particle} horizons) or parts
which, in principle, cannot be oberserved from our position. \\
An example for observations  n o t  satisfying $B_1$ is formed by the sample 
used for  gaining and calibrating spectra of Ia supernovae \cite{SeiSchwa2009}.\\

 - $B_2$ Observation time is long enough in order to guarantee
reliable data of cosmological relevance.\\
 
- $B_3$ Ambiguities in observation and theoretical interpretation (selection
 effects) are identified and taken into account by bias parameters.\\

An example for a bias parameter $b(z,k)$ is given by the expression for the
observable galaxy overdensity $\delta_g$ as a measure of the
underlying (average) matter density $\delta_m$:
$\delta_g=b(z,k)~\delta_m $ (\cite{Rassatal2008}, eq. (3)). 
It is unclear whether these demands on observation are satisfied, at present. 
In particular, selection bias concerning luminous objects may be
underestimated (\cite{TammSand2008}, p. 321).

But it is in observation that tremendous progress has been made in the past
two decades. 3-dimensional redshift surveys of
galaxies\footnote{redshift $z=\frac{\lambda'-\lambda}{\lambda}$ directly
  relates to distance $D$; for small distances, $z=H_0D$.} have been much 
extended. In
particular, this was done by the 2dF galaxy redshift survey (combined with the 
2QZ quasar redshift survey) (2003): patches of $2\times 2$ degrees have been 
probed and $221414$ galaxies ($23424$ quasars) measured out to $4\cdot 10^9~ 
$ lightyears (up to $z=0.22$) (2QZ: two $5 \times 75$ degree stripes both in the
northern and southern sky)\linebreak (http://www2.aao.gov.au/2dFGRS). Most 
impressive is the Sloan digital sky survey \cite{Eisensteinal2005},
\cite{Percivalal2007}: it comprises $\sim 10^6$ galaxies, with the subsample
of luminous red galaxies at a mean redshift $z= 0.35$ and 19 quasars at 
redshifts $z \geq 5.7$ up to $z=6.42$ (http://www.sdss.org). Cf. also the 
``Union Sample'' of Ia
supernovae containing 57 objects with redshifts $ 0.015 < z < 0.15 $, and 250
objects with high redshift \cite{Kowalskial2008}. In view of an assumed total of 
$\sim 10^{11}$ galaxies in the universe and the fact that angular position 
surveys extend only to depths of a fraction of the Hubble length, one cannot
say that these surveys are exhaustive. Moreover, in view of the fact that 
estimates of the mass-luminosity ratio lead to $\Omega_{lum}\simeq 0.005$ for
the relative density of luminous matter, the cosmological relevance of the 
galaxy surveys is questionable; they may amount only to a consistency
check. The scale 
of homogeneity for which averaging of the observed large structures 
(superclusters, voids) is reasonable, has steadily increased in the past and 
could grow further, in the future. At present, the size of the Great
  Wall, i.e., $\simeq 400~ Mpc$ seems to point to a homogeneity scale of 
$\geq 100~ Mpc$ \cite{Labinal2009}. The surveys described have been 
used to test homogeneity, e.g., by counts of luminous sources in a redshift 
range of $0.2 < z < 0.36$, albeit with distance calculations within the 
homogeneous and isotropic $\Lambda$CDM-model \cite{Hoggal2005}.

Isotropy with respect to our observing position also has been put to a 
test; a statistically significant violation of \linebreak isotropy for Ia 
supernovae at redshift $z<0.2$ and refering to deviation in the Hubble diagram 
(Northern and Southern Hemispheres) has been found \cite{SchwaWe2008}. Problems 
related to observations were investigated carefully by G. F. R. Ellis
\cite{Ellis1984}, \cite{Ellis2006}. 

\subsection{More on the standard cosmological model}
\label{subsection:morestandard}
In the standard model, the gravitational field and space-time are described by
a (pseudo-)Riemannian manifold with a homogeneous and isotropic Lorentz
metric. It is an expression of the Cosmological Principle, which
  alternatively can be formulated as (cf. section \ref{subsection:genhypo}): ``No matter particle (of the averaged out ideal cosmic
matter) has a prefered position or moves in a prefered direction in the
universe''. Consequently, the space sections of the spacetime manifold describing
the universe are homogeneous and isotropic in the sense of an average (on the
largest scales) over the observed matter distribution. The cosmological metric 
(gravitational potentials) is given by a Friedman-Lemaitre-Robertson-Walker 
solution 
(FLRW) of Einstein's field equations - with or without cosmological constant. 
The metric depends on a single free function $a(t)$ of cosmic time and allows 
for a choice among three space sections with {\em constant} 3-curvature
($k=0,~+1,~-1$). The parameter $k$ is related to the critical energy density 
$\rho_c = \frac{3c^4~H_0^2}{8\pi G}$ such that $k=0$ for $\rho =\rho_c$; $k>0$ 
for $\rho >\rho_c~$ and $k<0$ for $ \rho < \rho_c$. This follows from the Friedman
 equations. When formulated with dimensionless (energy-) density parameters 
$\Omega_x:=\frac{\rho_x}{\rho_c}$, where the index $x$ stands for $c$ 
(critical-), $d$ (dark-), $b$ (baryonic-), $t$ (total matter), respectively, and 
$\rho_{\Lambda}=\frac{\Lambda c^4}{8\pi G}$, $\rho_{k}=\frac{k c^4}{8\pi G
  a(t)^2}$, one of the two Friedman equations reads (trivially,
$\Omega_{c}=1$):\begin{equation}
   1=\Omega_t+\Omega_{\Lambda}+\Omega_k~\label{Fried}\end{equation} with
 $\Omega_t= \Omega_b+\Omega_d+\Omega_{radiation}$. Due to its smallness, we
 mostly will neglect $\Omega_{radiation} = \Omega_{\gamma}(1 + 0.2271
 N_{eff})$ with $\Omega_{\gamma}$ the photon density and $ N_{eff}$
 the (effective)  number of neutrino species (\cite{Komatsual2009}, p. 335).

The space sections for $k=+1$ are compact; those for $k=0, -1$ usally are called 
``open'' as if they could have only infinite volume. This misconception 
is perpetuated in otherwise excellent presentations of cosmology; in 
contradistinction, a sizeable number of space forms of negative curvature with 
finite volume were known to mathematicians since many years
(cf. \cite{Luminetal2003}, \cite{Ellis2007b}, p. 405). This is important because 
different topologies can be consistent with the WMAP-data \cite{Steiner2008}. 
   
The lumpiness of matter in the form of galaxies, clusters of galaxies, and 
superstructures is played down in favour of a continuum model of smeared out 
freely falling matter like in an ideal gas. Its particles follow timelike 
(or lightlike) geodesics of the FLRW-metric. Inhomogeneity then is
reintroduced through linear perturbation theory on this idealized
background. In two stages in the history of the universe, both with power-law 
expansion, the equation of state considered above refer to pressureless matter 
(baryon dominated universe) and to radiation where $p = 1/3~\rho$ (radiation 
dominated universe).\footnote{At redshift $z \sim 3600$, the period of 
matter domination follows the radiation-dominated one; decoupling of
  photons is set at $z \sim 1100$. For a detailed discussion of the standard 
model and the early universe cf. \cite{Boerner2003} or \cite{Mukhanov2005}.}
At present, a general equation of state $p=w\rho$, with $w$ being allowed to
be negative, is deemed necessary because the cosmological constant may be 
simulated by $p=-\rho$. 

Moreover, from observations alone, it seems unclear whether it is possible to
discriminate, in our neighborhood, between a Friedman model and spatially 
inhomogeneous models centered around our position and resembling a
  Friedman model (Lema$\hat{i}$tre-Tolman-Bondi- or Stephani exact
  solutions). For a review cf. section 2.3 of \cite{Celer2007}. Also, a metric
  combining the FLRW-model and ``a perturbed Newtonian setting'' has
  been used to approximately describe features of both the local
  universe and its large-scale structure \cite{BuElEl2009}.

The FLRW-metric describing the cosmological model does not care whether 
its primordial states are warm or cold. Only when the vanishing of the
divergence of the energy-momentum tensor of matter is interpreted as
describing the first law of non-relativistic thermodynamics, the expansion of 
the universe can be seen as an {\it adiabatic} process, with the ensuing
decline of temperature following the expansion of space. In consequence, it is 
possible to interpret the microwave background as a relic of an early, hot 
state of the universe. On the other hand, adiabaticity is violated at the end
of the inflationary period where particles and heat are generated. From 
local physical processes we expect the entropy of the universe to grow with
the expansion (deviation from homogeneity). In principle, statistical
mechanics (kinetic theory) is the only way for defining properly the concepts of 
temperature and entropy of the universe: no ``external'' heat bath is
available. Whether they make sense depends on the existence of an unambiguous
procedure for coarse graining in phase space. For the entropy concept,
cf. the point of view of a strong supporter (\cite{Penrose2005}, section 27).

Mathematically, the most important consequence of the FLRW-models
is that they show the occurence of infinite density - as well as a metrical
{\it singularity} appearing in the {\it finite} past: the famous big
bang. By mathematical theorems of Penrose and Hawking \cite{HawkingEllis1973}, 
singularities receive a {\it generic} geometric significance within 
cosmological model building. Their physical aspects were studied by 
Belinskii \& Khalatnikov (\cite{BelKhalat1969}, \cite{BelKhalatLif1972}).
From the point of view of observational cosmology, the infinities connected
with the big bang cannot and need not be taken seriously.

We have seen in section \ref{subsubsection:peculiar} that the ``predictive''
 power of the standard cosmological model is nothing more than an expression of
self-consistency. By use of the cosmological model, from the temperature at 
one {\it past} time, e.g., at the decoupling of radiation and matter
$T_{dec}$, the present background photon temperature would be calculated 
to be $T_{phot}(0)=\frac{T_{dec}}{1+z}$ and the baryon temperature $
 T_{bary}(0)=\frac{T_{dec}}{(1+z)^2}$. The temperature of the neutrino
 background then is fixed. The consistency problem comes up because $T_{dec}$
 can be calculated via the Saha equation which includes $\eta =
 n_B/n_{\gamma}$, a number which can be read off from the CMB. Of course,
 this single chain of arguments is supported consistently by others; e.g., the 
fluctuations in mass density at
decoupling must be such that their growth (gravitational instability) until 
now is consistent with the observed relative anisotropies of $10^{-5}$ in the 
otherwise isotropic CMB etc. As in other parts of physics, there is a net 
of theoretical conclusions relating empirical data and theory.

The standard cosmological model faced the task of getting away from 
the homogeneity and isotropy of the averaged out large scale matter content 
in order to arrive at an explanation of the large scale structures consistent 
with the required time periods. The hypothesis of primordial adiabatic Gau{\ss}ian
density fluctuations with a nearly scale-invariant spectrum together with 
various competing scenarios as {\it cold} or {\it hot dark matter} (in the
form of weakly interacting particles), {\it cold baryon matter}, cosmic string 
perturbations, local explosions etc, for some years had not been consistent 
with the full range of extragalactic phenomena \cite{Silk1987},
\cite{PeeblesSilk1990}, {\cite{Bothun1998}. By now, this debate seems to be 
ended: the cold dark matter scenario is accepted.

\subsection{The concordance model of the universe ($\Lambda$CDM)}
\label{subsection:concord}
Due to the observations pointing to an {\em accelerated} expansion\footnote{
  The so-called {\em deceleration} parameter is defined by $q= - \frac{a
  \ddot{a}}{\dot{a}^2}.$ Negative $q$ means acceleration.} of the 
universe in the present era, and due to much progress in astrophysical structure
formation theory, the standard cosmological model of the early 90s took the 
following turn: (1) In structure formation, cold dark matter, i.e.,
non-relativistic particles subject to gravity, and able to contribute to the 
growth of matter inhomogeneities (against radiation drag) better than and before 
baryons can do so, came to play a decisive role; (2) the space sections of the
FLRW cosmological model were assumed to be flat ($k=0$); (3) the cosmological 
constant $\Lambda \neq 0$ mimicking a constant energy density became 
re-installed. A consequence was that due to $\Omega_k=0$ in the Friedman 
equation (\ref{Fried}): $\Omega_t + \Omega_{\Lambda} = 1.$ Because 
$\Omega_t$ contains both, baryonic and dark matter, and due to 
$\Omega_m=\Omega_{b}+\Omega_{d}\simeq 0.25$, a missing mass 
$\Omega_{\Lambda}\simeq 0.75$ resulted, named ``dark energy''
  \cite{Turner1998}. This naming occured due to the original interpretation of 
the cosmological constant as a representation of ``vaccum energy'' in the
  sense of the energy of fluctuations of quantum fields (cf. the end of 
\ref{philoprob}). 

Observation of the luminous-galaxy large-scale-structure also showing
baryonic acoustic oscillations (BAO), of the temperature anisotropies of the 
cosmic backgound radiation (CMB) as well as the determination of the value of 
the Hubble constant, and the age of the universe, all have been used to
support the $\Lambda$CDM model. In particular, CMB measurements by the WMAP 
(Wilkinson Microwave Anisotropy Probe)-satellite as reflected in the acoustic 
peaks from baryonic and dark matter give information on (\cite{WMAP2008}, 
table 7, p. 45):

 - the  geometry of space sections ($\rightarrow k$ small, $ -0.0179<\Omega_k
  < 0.0081$);\\
 - matter energy density $\Omega_m =\Omega_{b} +\Omega_{d}\sim 0.258\pm 0.03 $;\\
 - vacuum energy density $\Omega_{\Lambda} \sim 0.726 \pm 0.015 $;\\
 - baryon density $\Omega_b \sim 0.0456 \pm 0.0015$;\\ 
as well as about further cosmological parameters:\\
 -  cold dark matter density $\Omega_d = 0.228 \pm 0.013$;\\ 
 - tilt $n=0.960 \pm 0.013$ of the initial power spectrum $P_{initial} \sim
 \bar{k}^n$ where $\bar{k}$ is the wave number of the initial fluctuations, 
\footnote{In fact, the amplitude of curvature fluctuations
  is defined by $\Delta_{\cal R}(\bar{k})^2:= \Delta_{\cal R}(\bar{k_0})^2
  (\frac{\bar{k}}{\bar{k_0}})^{n(\bar{k_0})
  -1+\frac{1}{2}\frac{dn}{dln(\bar{k})}}$ if $n$ is allowed to
  vary. $\bar{k}_0=0.002~Mpc^{-1}$.}\\  
 - the Hubble constant $H_0= 70.5 \pm 1.3~ kms^{-1}(Mpc)^{-1}$.

All these results are based on the CDM model for structure formation.
Two further numbers $w_0, w_z$ parametrize a generalized equation of state 
$p=w(z)\rho$, with $w(z)= w_0 +\frac{z}{1+z}w_z $ being allowed to become 
redshift-dependent \cite{Linder2003}. A ``minimal'' parameter base of the
  $\Lambda$CDM model is given by $\Omega_m, \Omega_c,\Omega_{\Lambda}, \tau, 
\Delta_{\cal R}^2, n$ where $\tau = 0.084 \pm 0.016 $ is the optical depth due 
to reionization (electron scattering) \cite{Komatsual2009}. A 7-parameter
  model with $\Omega_m, \Omega_b,\Omega_d, w_0, w_a, h, n$ is considered in 
\cite{Rassatal2008}. The errors in $\Omega_{d}$, $\Omega_{b}$, and the
  Hubble constant are claimed to be 3\% (\cite{WMAP2008}, p. 2-3). From WMAP,
  the baryon accoustic peaks and supernovae, a bound on the summed neutrino
  masses $m_{\nu}$ (of the standard model of elementary particles) has been 
deduced: $\Sigma m_{\nu}\leq0.62~ eV$ \cite{GoHanal2006}. Eventually, this
will be confronted with precise measurements of the neutrino masses on Earth.

\subsection{Matter content of unknown origin}
\label{subsection:unobserved}
\subsubsection{Dark matter}
\label{subsubsection:darkmat}
From observation of the bulk motion of galaxies and clusters of galaxies 
in the past 65 years, it is known that more mass than that of the luminous 
objects must be present. This is needed for an understanding of the dynamics
of such objects, for galaxy formation, and for the interpretation of the
results of weak gravitational lensing from clusters of galaxies. The
mass is missing in and around galaxies (halos). For a review cf. 
\cite{Roos2010}. As we know from section \ref{subsection:concord}, baryons, 
mostly in the form of gas, contribute to only ca. 4\%-5\% of the relative 
critical density $\Omega_c=1$ (\cite{LiddleLoveday2008}, p. 90). Besides being 
required to provide an enhancement of gravity, dark matter is assumed to be 
``non-interacting'', i.e., pressureless, otherwise. Computer
simulations like the Aquarius Project \cite{Springal2008} or MS-II have 
excellently taken into account and reproduced dark matter: ``from halos
similar to those hosting Local Group dwarf spheroidal galaxies to halos 
corresponding to the richest galaxy clusters'' (\cite{BoylanKolchinal2009}, 
abstract).

For a tentative explanation of dark matter either new cold (i. e., 
non-relativistic) particles (WIMPs,\footnote{Weakly interacting particles.} 
axions, neutralinos or other light supersymmetric particles, primordial black 
holes), as well as Q-balls, and other unobserved exotic objects were 
suggested. The composition of dark matter particles is closely bound to 
baryogenesis \cite{Buchmueller2007}. Eventually, dark matter particles must
  be found in accelerator-experiments, and their masses measured, in order that
  their existence be more than speculative. Alternatively, new theories of 
gravitation have been suggested removing the need for dark matter, as
are Modified Newtonian Dynamics (MOND) (cf. {\cite{SandMcGaugh2002},
Scalar-vector-tensor-gravity (STVG) \cite{BrownMoffat2006}, translational 
gauge theory \cite{HehlMashoon2009a}, \cite{HehlMashoon2009b}, etc. Up to now, 
none of the particles invoked were seen, and none of the alternative theories 
were able to replace Newtonian theory in all aspects. From the modeling of galaxy
formation, hot dark matter in the form of neutrinos seems to be excluded.

\subsubsection{Dark energy}
\label{subsubsection:darkerg}
Since about a decade, observation of the luminosity-redshift relation of
type Ia supernovae has been interpreted as pointing to an {\em accelerated} 
expansion of the cosmos \cite{Riessal1998}, \cite{Perlmutter1999}. The simplest
explanation is provided by a non-vanishing cosmological constant
$\Lambda$ within the standard cosmological model. In this case, dark
energy would be distributed evenly everywhere in the cosmos. It
apparently has not played a significant role at early times although
reliable knowledge beyond $z=1$ is not available 
(\cite{CaldwellKamionkowski2009}, p. 8).

Besides the cosmological constant, tentative dynamical explanations have 
been given for cosmic acceleration. There, the main divide is between those 
keeping Einstein gravity or proposing alternative theories. In the first
group, we find, on the matter side,\\ 

- a new scalar field $\Phi$, named {\em quintessence}. Strictly 
speaking,\linebreak ``quintessence'' stands for a number of model theories for the
scalar field like cosmic inflation stands for a large 
number of different models.\footnote{In a specific model, the scalar field has 
been named ``cosmon'' \cite{Wetterich2002}. Another suggestion leads to a
pseudo-Nambu-Goldstome boson \cite{Friemanal1995}.} Quintessence models 
work with an equation of state $w=\frac{p}{\rho}$ with $-1 < w <
-\frac{1}{3}$. The kinetic energy term is the usual $\nabla_i\Phi
\nabla^i\Phi$ while for an extended set of models, i.e., k-essence theories,
the kinetic term may read as $f(\nabla_i\Phi \nabla^i\Phi)~ g(\Phi)$ with 
arbitrary functions $f,g$. In both sets of theories, the scalar field can
interact with baryonic and/or dark matter. There are even more speculative
approaches taking the kinetic energy terms to be {\em negative} (phantom
fields) \cite{CaKaWe2003}. For further alternative theories of
gravitation, cf. the reviews about the understanding and consequences of cosmic
acceleration by \cite{Celer2007}, \cite{SilvestriTrodden2009} and
\cite{CaldwellKamionkowski2009}. Within Einstein gravity, another road has
also been taken:\\

 -  By a suitable averaging procedure. It is argued that the differences in 
gravity between observers in bound systems (e.g., galaxies), 
and volume-averaged comoving locations within voids (underdense regions) in 
expanding space can be so large as to significantly affect the parameters of 
the effective homogeneous and isotropic cosmological model \cite{Wiltshire2007}.
A great deal of research is available \cite{Buchert2000},
\cite{Buchert2008}, and has lead to testable consequences
\cite{LiSchwa2007}. The observations seem not yet conclusive with regard to
  whether we are located in such an underdense void of an extension
  $200-300~Mpc$ \cite{HuSar2009}.\\

If we refrain from accepting proposed ad-hoc-changes of the Friedman
equations, among the theories suggested as replacements of Einstein gravity 
there are theories with higher-order field equations.\footnote{That is, with 
Lagrangians of higher-order in the curvature tensor.} In one 
class, the curvature scalar $R$ is replaced by an arbitrary function $f(R)$. For a
general review cf. \cite{SotiriouFaraoni2008}; for a critical status 
report {\cite{Straumann2008}. Again, scalar-vector-tensor theories of
  gravitation and vector-tensor theories \cite{BalDehn2009} were put
  forward. In ``make-believe cosmology'' models with a higher number
  of spacelike dimensions are considered, e.g., five-dimensional
  braneworld models and also string related theories. Cf. section 
\ref{subsection:multiverse}. 

In comparison with dark matter, the observational status of dark energy
remains less secure. Observed is a dimming of the luminosities of type Ia
  supernovae from the luminosity-distance relationship. Together with the
  homogeneity assumption this leads to acceleration (\cite{Celer2007},
  p. 17, \cite{SeiSchwa2008}).
  With further assumptions added, e.g., of flat space sections, dark energy
  then is reached. At present, the only promising method for its
  future empirical grounding seems to be (statistical) weak
  lensing. In contrast, for dark matter the case is very strong, cf. \cite{McGaughal2007}, \cite{Roos2010}. 

\subsection{Further conceptual pecularities of the standard model}
\label{philoprob}
As discussed in section \ref{section:content}, the standard model of cosmology 
is not free from epistemological and methodological problems. To list one
more: Newton's absolute space appears in disguise in the form of an {\it 
absolute reference system}. In particular, (absolute) {\it cosmic time} or era 
is without {\it operational} background: the only clock measuring it is the 
universe itself. {\it By definition}, cosmic time is identified with atomic 
time. By what sequence of clocks the measured time intervals of which must be 
overlapping, can precise time keeping be realized for the full age of the 
universe? In particular, which ``clocks'' to use before structure formation, 
before nucleosynthesis, before baryogenesis, during the inflationary phase?
From the radiocarbon method we know that ``radiocarbon years'' must be 
recalibrated to correspond to ``calendar years''. Such a re-calibration (in
terms of radioactivity- and astronomical clocks etc) is necessary also for
cosmological time. In the very early universe described by quantum cosmology, 
only some sort of ``internal'' time seems to be possible. 

Also, there is {\it no operational} way of introducing simultaneity. The 
local method of signaling with light cannot be carried out, in practice, if 
distances of millions of light years are involved and the geometry in between 
the large masses is uncertain. It cannot be used, {\it in principle}, for the 
full volume of space if event horizons are present. The cosmological models 
containing the concept of ``simultaneous being of part of the universe'' 
(technically, the space sections or 3-spaces of equal times) are catering to 
past pre-relativistic needs. For the relativistic space-time concept, access 
to the universe is gained through the totality of events on
and within our past light cone. Hence, ``simultaneous being'' must be
replaced by ``what may be experienced at an instant at one place'' (a stacking 
of light cones). Some of the objects at the sky, the radiation of which we 
observe today, may not exist anymore. 

A special case of the {\em hierarchy problem}, i.e., the so-called {\em 
cosmological constant problem}, arises if the cosmological constant $\Lambda$ is
not seen as just an additional parameter of classical gravity, but
interpreted as the contribution by vacuum fluctuations of quantum field
theory. In this case, its value should be immensely larger than the
value derived from observations by a factor of $\sim 10^{60}$ (in
theories with supersymmetry), or $\sim 10^{120}$ (no supersymmetry). In
\cite{Smolin2009} a solution to this problem within quantum gravity has been
suggested. 

\section{The inflationary flash}
\label{section:inflafla}

\subsection{Particle cosmology}
\label{subsection:particlecos}
As we are going back in cosmological time, a remark concerning particle
cosmology seems in order. While the temperature of the universe heats up toward 
the big bang, it is assumed that matter undergoes a number of phase
transitions. All those happening before the so-called electroweak phase 
transition at $\sim (100-200)~ GeV$, occur at energies not yet attainable in 
the laboratory (accelerator particle physics). All are speculative, as e.g., 
the grand unification phase transition at which the strong interaction unifies 
with the weak and electromagnetic forces. The confinement-deconfinement
  (QCD)-phase transition at $10^{-5}s$ after the big bang seems to be the only 
one in future reach of accelerator physics. Cosmic inflation preceeds all the 
mentioned events; whether it is ending in a phase transition or not, is
debated. After the end of inflation, copious particle creation and then
thermalization is asumed to occur followed by baryogenesis. Cosmological 
modeling after inflation is characterized by a change of paradigm when 
compared to later eras: while, in principle, the description of matter by a 
continuous distribution is retained, in practice matter is differentiated into
elementary units: elementary particles, nuclei, atoms and their reactions;
they interact, can be produced or anihilated. The interplay of elementary 
particle reaction rates and the expansion rate of the universe requires 
different equations of state for different particle species at the same
epoch. Nuclear physics comes in much later: the end of primordial 
nucleosynthesis is assumed to have happened at $\simeq 10^2 s$ after the big 
bang. Particle physicists are interested in the very early universe as a
testbed for their theories concerning high energies. While by the later evolution 
of the cosmos limits  are set on such theories (from CBM), the direct
contributions of elementary particle physics to the early universe are 
speculative.

Again, cosmological modeling of the early states of the universe is based on a
number of hypotheses, simplifying the modeling. A selection would be:\\

 - $C_1$ Baryogenesis occurs after the end of inflation.\\
 
As to $C_1$, the end of inflation (reheating) is not well understood;
it is difficult to reconcile the slow-roll conditions with the known
couplings of particle physics candidates for the inflaton. The origin of the 
matter-antimatter asymmetry in the cosmos must and can be explained (cf. 
\cite{Buchmueller2007}). A number of theories have been suggested, some of
them using leptogenesis (sphaleron-interaction) \cite{Pilaf2009}.\\

 - $C_2$ Both, the reactions and reaction rates of individual particles, and
   collective phenomena are important in the early universe.\\

The assumed occurence of phase transitions cannot be understood without taking
into account collective interactions. For a review of such phase transitions
in the early universe cf. \cite{BoVeSchwa2006}.\\

 - $C_3$ Elementary particles do not interact gravitationally;
   gravitation acts merely as an external field.\\

This assumption expresses the subordinate role gravitation plays in the
modeling of the early universe despite the assumption that then matter was 
extremely condensed. The gravitational field is assumed to show up only in the 
expansion of the universe or, perhaps, in pair production of elementary 
particles, if quantum field theory in curved space as we understand it is 
applicable (there exists not yet a fully worked out model for strong
curvature). For special aspects cf. \cite{HollandsWald2002a},
\cite{HollandsWald2002b}, \cite{BuchholzSchlemmer2007}. \footnote{Of course,
  in the very early universe, the gravitational field might not exist on its 
own but be united with the other fundamental interactions in a Super Grand 
Unified Field.}\\ 

 - $C_4$ Temperature and entropy of the universe are well defined after
   (local) thermodynamic equilibrium is reached.\\

 - $C_5$ While, in epochs after inflation, matter is in thermodynamical
   equilibrium, different particle species can and will decouple from
   the equilibrium distribution.

As to the application of thermodynamics and kinetic theory to the early
universe ($C_4,~C_5$), it is known that, in the FLRW cosmological models, an 
exact equilibrium distribution is permitted only in two limiting
cases: the ideal radiative model (rest mass of particles is zero) and
the ``heavy mass''-model (infinite rest mass) \cite{Bernstein1988}. 
Thermodynamically, the expanding universe is treated as a {\it
  quasi-static} system with a relaxation time small with regard to the 
expansion (Hubble) time.\footnote{Relaxation time, for massive
  particles, is related to mass diffusion or heat transport etc. For
  massless particles it may be approximated by the collision time and
  does not depend on volume.} This is called {\em local
  thermodynamical equilibrium}. Whether such a concept can be valid for 
{\em infinite} volume (open space-sections with $k=0, -1$) seems
questionable. From this perspective, a ``small'' universe 
would be preferable. The time dependence of cosmic temperature implied by the 
cosmological model (adiabatic cooling), could be interpreted as a 
characteristic sign for the universe being a {\it non-equilibrium} system. 

\subsection{The inflationary model}
\label{subsection:inflation}
If the validity of the FLRW-cosmological models is extrapolated
to very early epochs, an inflationary period between $\simeq 10^{-36}s$ and 
$\simeq 10^{-34}s$ after the big bang is assumed to have happened. During it, 
all distance scales in the universe must increase by at least 75 e-folds 
(\cite{Mukhanov2005}, p. 239). In connection with the cosmological standard 
model, a number of questions then could be answered:\\
 - What makes the universe as isotropic and homogeneous as it
is (horizon problem)?
 - Why does the overall density parameter $\Omega$ differ from
   $\Omega_{\mbox{c}}=1$ by only by very little (flatness problem)?
 - How can the ratio $\eta =~\frac{\eta_B}{\eta_{\gamma}}~\simeq
(4-7) \cdot 10^{-10}$ be explained (entropy problem)?

In order to answer these questions, the idea of the {\it inflationary
  scenario} was
invented \cite{Guth1981}, \cite{AlbrechtSteinhardt1982}, \cite{Linde1990}, 
\cite{KolbTurner1990}. Its characteristic feature is a scalar field $\phi$, 
the ``inflaton''\footnote{More precisely, the inflaton is the field quantum 
of the inflaton field.}, which is supposed to dominate the matter
content at very early epochs. This scalar field must be very weakly coupled to 
all other matter fields. Usually, although not necessarily, $\phi$ is taken to 
be the order parameter of a phase transition from a symmetric phase with high 
energy corresponding to $\phi = 0$ (false vacuum) to a phase with {\it broken
  symmetry} and $\phi =~\mbox{const} \neq 0$ (true vacuum). An analogue would 
be the delayed transition from the gaseous to the fluid state with
undercooling. The phase transition is made to start at $\simeq 10^{-35}$
seconds after the big bang. Dynamically, it is tripartite: after the tunneling 
of a potential barrier between the false and the true vacuum, a slow
descent (``role-down'') toward the true vacuum (supercooling) to a
period of field oscillations, (reheating) must occur. In this last
interval, the inflaton 
decays into the matter particles/fields we see today, and by producing heat. 
The reheating process is non-adiabatic and claimed to bring an increase in the 
entropy (of the universe) by a factor of $10^{130}$. \footnote{During the 
inflationary phase, entropy grows linearly with cosmic time $t$, afterwards 
only with $ln~t$ (\cite{Kiefer2007}, p. 319).} The equation of state of the 
inflaton field is unusual if compared with materials in the laboratory: its 
pressure is negative with $p=-\rho$ ($w=-1$). Gravitational attraction is 
overwhelmed by repulsion responsible for the rapid expansion of the universe 
during the inflationary period. 

A reason behind the many inflationary models is the ambiguity in
potential energy of the inflaton field: it may be taylored at will. In
some of the models investigated by now, the phase transition is
pictured as a nucleation of bubbles of the
broken-symmetry phase within a matrix of the symmetric phase. During
supercooling such a bubble can grow exponentially by 40 - 50 orders
of magnitude (of 10) and more within a time of the order of a (few
hundred) $\cdot 10^{-35}$  seconds. The gravitational field during
the exponential growth is described by de Sitter's solution of the
field equations (with constant Hubble parameter), the space sections of which are 
{\it flat} ($k=0$). By construction, the inflationary model
can solve both the entropy and the horizon 
problems: the presently observable part of the universe lies within a single 
inflating bubble. This means that, at the epoch of decoupling of photons and 
baryons, the various regions of the universe from which the cosmic microwave 
background originated have been causally connected. The model is said to also 
remove the flatness problem: inflation drives the density prameter $\Omega$
toward one \cite{Ellis1991}. Whether $\Omega = 1$ is desirable or not, seems to 
be entirely up to one's private beliefs, though.\footnote{$\Omega = 1$ is an 
unstable fixpoint in the phase diagram of the time evolution of the
Friedman models.} There are also inflationary models with negative and 
positive 3-curvature $k$ \cite{Bucheral1995}, \cite{Gott1986}. Hence,
it seems questionable whether ``the flatness of the universe'' is an
unavoidable consequence of inflation (\cite{Mukhanov2005},
p. 354).\footnote{Also, as noted by R. Penrose, if theory implies flat space 
sections, no observation, as small as its error bar can be made, will be able 
to exlude nonzero curvature (\cite{Penrose2005}, p. 772).} We note that
the inflaton field might be inhomogeneous and yet not violating the
homogeneity and isotropy of the energy-momentum tensor of the cosmological
model; the overall homogeneity would then be lost, however. 

Although debates about the inflationary model have not ended
(cf. {\cite{AlbrechtSorbo2004}, \cite{HollandsWald2002a}, \cite{Kofmanal2002},
  \cite{Turok2002}, by the following result its acceptance became
  overwhelming: through quantum fluctuations of the inflaton field, the model 
was able to provide the nearly scale invariant spectrum in the growing mode of 
(adiabatic) density perturbations which had been required from
observations.\footnote{ An admixture of isocurvature (non-adiabatic)
  perturbations below 10\% (3\%) seems to be permitted \cite{Trotta2007}, \cite{ValGian2009}.} To make the amplitudes fit the density fluctuations reflected by 
the anisotropy of CMB, fine-tuning is required, though. In this context, it
has been shown that large-angle (low-$l$) correlations of the CMB
(from the 3-year WMAP-data) exhibit statistically significant
anomalies. This is weakening ``the agreement of the observations with
the predictions of generic inflation'' (\cite{CoHuSchwaSta2007}, p. 16).

\subsection{$\Lambda$CDM-questionaire (implying inflation)}
\label{subsection:concoquestion}

While the inflationary model needed for the $\Lambda$CDM model has solved a 
number of problems, it created others:\\
 
\noindent - By what physics are the initial conditions for inflation generated?\\
 - What is the inflaton field?\\ 
 - What is tested by present observations: the nearly scale-invariant 
   spectrum of density perturbations, or the inflationary scenario, in toto?\\ 
 - What is dark energy?\\ 
 - Why dark energy has become dominant only ``recently'' in the evolution
   of the universe (coincidence problem)?\\
 - Did dark energy play a role in the formation of large scale 
   structure, or not?\\
 - Is an interaction of dark matter and dark energy excluded?\\

At present, there seems to be no consent on a fundamental theory for
the very early universe in which the inflationary model is embedded and its 
initial conditions fixed. Cf. critical remarks by \cite{Penrose1989b}.
\footnote{Cf. however \cite{Bojowald2002} with a worked out suggestion that 
quantum geometry lead to inflation.} The inflaton is not the 
Higgs particle (both are not observed). Is it connected to a model of hybrid
inflation (2 scalar fields!) with the s-neutrino as the inflaton?
\cite{Antuschal2005} Is there a link to the scalar field introduced in a later 
epoch and named ``quintessence'' (Cf. section \ref{subsubsection:darkerg}). Will
there be a {\it technically accomplished} model for inflation still lacking?
\footnote{For different inflationary models including chaotic, double, hybrid, 
new and eternal inflation cf. \cite{Guth1997}, \cite{LiddleLyth2000}.} What
determines the high energy of the false vacuum? What kind of traces of 
the inflationary period can we {\it observe}? One such effect following from 
inflationary models is a stochastic background of primordial gravitational 
waves: metric tensor modes could be seen in the polarization measurements of 
CMB. So far, they have not (yet) been detected. If observed, certain 
inflationary models with respect to others could be ruled out. If not found, 
this also can be reproduced by some models. Gravitational waves from inflation 
are not to be mixed up with ``gravitons'' eventually generated during the 
Planck era, nor with the still different ``gravitons'' claimed by string
theory. 

The coincidence problem is alleviated if cosmic acceleration is
modeled by space- and time-dependent fields replacing the cosmological
constant; a fine-tuning of their contribution to the energy density needed can 
always be made such that it is largest late in the evolution of the universe.
In view of the merely indirect empirical tests through consistency of the full 
cosmological model, the inflationary scenario is still rather speculative.

\section{Originative cosmology}
\label{section:extrapol}

\subsection{Quantum gravity}
\label{subsection:quantgrav}
In a third stage of cosmological modeling, the epoch around and before the 
Planck time ($10^{-44}~s$) is briefly dealt with. At such extremely
early epochs, quantum mechanics and quantum field theory are
applied. At present, a
consistent and mathematically rigorous quantum field theory of gravitation,
i.e., {\em quantum gravity}, is under construction but still not
completed.\footnote{This is no surprise, when we think that even quantum field 
theory in Minkowski space has not yet been made mathematically rigorous in all 
aspects.} Nevertheless, within general relativity, intriguing schemes like 
{\em canonical quantization} in the geometrodynamics approach \cite{DeWitt1967}, 
\cite{Wheeler1968}, \cite{Kuchar1973}, its gauge theoretical variant {\em loop 
quantization} \cite{Ashtekar2009},\cite{Ashtekar2007}, \cite{Thiemann2007}, 
\cite{Rovelli2008}, {\em covariant quantization}, e.g., in the form of Feynman 
path integral quantization \cite{Hamber2009}, and the (numerically
implemented) models of causal dynamical triangulation  \cite{Loll1998}, 
\cite{Ambjornal2005} are pursued with impressive success.\footnote{It is
    an open question whether these different approaches will lead to equivalent 
quantum theories of gravitation.}   Some general hypotheses are made:\\

 - The gravitational field must be quantized around and before the Planck epoch.\\
 - Unlike in the procedure for other fields, quantization of gravity must
   be done in a background independent manner (in canonical quantization).\\
 - All local and global degrees of freedom of the gravitational field must be 
   taken into account.\\
 - Einstein's field equations hold right up to the big bang singularity.\\

That gravity ought to be be quantized is the majority vote. Some think that 
quantization must be performed within a theory in which all fundamental 
interactions are united, e.g., a claim made by string theory. At present, 
string theory does not yet noticeably contribute to a solution of the most 
pressing questions in quantum gravity; it still is in ``a rather preliminary 
stage'' (\cite{BlauThe2009}, p. 753). Few believe in gravity as a 
{\it classical} field generated, perhaps, as an {\it effective} field by the 
other fundamental interactions.\footnote{This is not to be mixed up with
  gravity dealt with as an effective quantum field theory with a high-energy 
cut-off.} Looked at from usual field quantization, at the 
root of the difficulties with quantization of gravity is its (perturbative) 
non-renormalizability. From a more technical point of view, quantization with
(Hamiltonian and diffeomorphism) constraints, as in the case of the 
Hamiltonian formulation of general relativity, is a hurdle. Moreover, it is
not entirely clear whether it suffices to quantize the 
gravitational field on a continuous space-time or, whether the very concept of 
a manifold ought to be replaced by discrete sets (causal set theory), 
combinatorily defined discrete structures like graphs, or spin networks (cf. 
\cite{Smolin2005}, \cite{Sorkin2005}). In loop gravity, while continuous
3-geometries still are investigated, area- and volume operators with a discrete 
spectrum do appear. Whether they are observables in the usual sense, i.e.,
  commuting with the diffeomorphism constraints, is not entirely
clear.\footnote{For a detailed discussion of the volume operator cf. 
\cite{Thiemann2007}, Secs. 13.1-13.6, pp. 432-457.} Background independence 
means that quantization should not rely on a metrical
structure but, at most, on a differentiable manifold 
(cf.\cite{AshtekarLewandowski2004} \cite{Giulini2007}. Consequently, a
lot of advanced mathematics is required. As no empirical input is
available at present, ``mathematical consistency is the
only guiding principle to construct the theory'' (\cite{Thiemann2007},
p. XX). The recent endeavour to derive rigorous results belongs into mathematical
physics. For  a critical discussion cf. \cite{Nicolaial2005},
\cite{Nicolaial2006}. Quantum gravity is said to apply to two main systems: the 
very early universe (quantum cosmology) and to evaporating black holes. 

\subsection{Quantum cosmology}
\label{subsection:quantcos}

\subsubsection{Law of initial conditions?}
On the one hand, application of quantum mechanics to the universe is seen as an 
{\it intermediate} step in between the big bang and the inflationary epoch
with the aim of providing initial conditions for inflation. But quantum 
cosmology also has been taken as a program for a {\it cosmogonic} theory: an 
attempt to construct a theory {\it determining uniquely} the initial
conditions of the universe \cite{HartleHawking1983},
\cite{Vilenkin1988}, \cite{Gell-MannHartle1990}. Turned around: as a program
for a theory avoiding the big bang singularity.\footnote{From quantum
  cosmology, we may expect more than forming a ``toy model for full quantum 
gravity in which the mathematical difficulties disappear'', cf. 
(\cite{Kiefer2009}, p. 894).} Such an endeavor makes sense only if the
universe itself 
carries the rationale for its initial data. If transferred to human life, this 
would mean that the reason for us coming to life does not lie in our parents 
but in ourselves. Strange as this thought may be (above the level of
protozoans): a human being and the universe are quite different systems. It 
seems plausible, philosophically, that the cosmos cannot be thought of without 
the inclusion of a reason for its coming into being. In classical theory, the 
very idea of {\it prescribing uniquely} the initial data of a system by help
of its {\em dynamics} is violating the spirit of physics. Perhaps, quantum 
theory could make the difference. For a positive suggestion in this direction 
within quantum cosmology, cf. \cite{Bojowald2001}, \cite{Bojowald2003}.  

\subsubsection{The Wheeler-DeWitt equation}
In the Hamiltonian formulation, space-time is foliated into space sections, 
and the Einstein field equations are decomposed into time-evolution equations and 
constraint equations on the 3-geometries $^3 g$. Canonical quantization leads 
to the Wheeler-DeWitt equation (WDW) for the {\it wave function} of the universe 
$\psi$, a formal analogue of the stationary Schr\"odinger equation \footnote{In
  reality, WDW comprises an infinite number of equations.}. It is a functional
$\psi[^3 g, \phi]$ of the geometry of space sections and the matter fields 
$\phi$ and hence defined on an infinite-dimensional space called 
{\it superspace}. The spacetime geometry can be pictured as a trajectory in 
superspace. The wave function of the universe represents the superposition of 
all possible space-time geometries correlated with matter functions
\cite{Zeh1986}. It is assumed to be a pure state. Mathematically, the 
WDW-equation is not well defined (factor ordering and regularization
problems). Nevertheless, one of the successes of the canonical approach is
that its semiclassical approximation bridges the gap to quantum field theory
on a fixed background \cite{Kiefer2009}.
 
In model calculations, isotropy and homogeneity of the space geometry is 
assumed and leads to a wave function $\psi$ depending on just one geometric 
variable: the scale factor $a$ of the Friedman models. Usually, only a single 
scalar matter field $\phi$ is taken into account such that $\psi = \psi[a, 
\phi]$. In this case, the infinite dimensional superspace is reduced to a
finite number of degrees of freedom, i.e. to {\it minisuperspace}. 

Despite this technical simplification, the main problem cannot be circumnavigated:
a {\it unique} solution of the Wheeler-DeWitt equation is obtained only if a 
{\it boundary} condition for $\psi$ is chosen. Several suggestions to this end 
have been made. In the path integral formulation \cite{HartleHawking1983},
\cite{Hawking1984} $\psi$ is determined by a summing over all paths describing 
{\it compact} {\em euclidean} 4-geometries with regular matter fields. All
4-geometries must have a given 3-geometry as their boundary 
(no-boundary-condition)\footnote{Cf. C.J. Isham \cite{Isham1987}: ``the
universe is created ex nihilo since the 4-manifold has only the connected 
3-space as its boundary''.}. An alternative condition is Vilenkin's quantum 
tunneling from nothing (where ``nothing'' corresponds to the vanishing of the 
scale factor a): the universe is nucleating spontaneously as a DeSitter space 
\cite{Vilenkin1982}, \cite{Vilenkin1986}. This boundary condition has been 
criticized on the ground that it equally well describe tunneling {\it into} 
nothing. For a detailed discussion cf. (\cite{Kiefer2007}, section 8.3, 
{\cite{Kiefer2009}, section 4.2). In loop quantum cosmology, the 
WDW-equation is replaced by a discrete evolution equation.

Because the dynamical equations follow from the constraints on the
spatial hypersurfaces}, the wave function of the universe cannot depend 
on an {\it external} time parameter as is cosmic time. In minisuperspace, the 
Wheeler-DeWitt equation is a {\it hyperbolic} differential equation the
dynamics of which is depending on two variables, $a$ and $\phi$, both of which 
can play the r$\hat{o}$le of an {\it internal} time. The ambiguity in the 
selection  of an internal time parameter permits reinterpretation of the
WDW-equation as a Klein-Gordon equation. In particular (cosmological)
  models, the (bounded) volume of the space sections are used as a measure of 
time. At the big bang, in loop quantum gravity, the (degenerated) eigenvalue
  of the volume operator is zero.

\subsubsection{Puzzles of quantum cosmology}
An acceptable quantum cosmology will have to solve three internal problems:\\ 
- to give a definition of time,\\ - to determine the role of ``observers'',\\ 
- to describe the ``emergence'' of a classical universe from the quantum one,\\
plus one external:\\  - to link quantum cosmology with empirical data.\\

The striking inequality in the treatment of time and space is an inheritance from
non-relativistic quantum mechanics. Presently, at best, time appears as a
notion in a semiclassical approximation scheme (\cite{Kiefer2007}, section
5.2). For a detailed discussion of the ``quantum problem of time''
cf. (\cite{Thiemann2007}, section 2.4).\footnote{It has also been argued that 
time can be eliminated altogether \cite{Barbour1993}.} 

A straightforward application of the Copenhagen-interpretation of quantum 
mechanics to the wave function of the universe does not make sense. 
Who is the classical observer carrying out preparation- and other
measurements? A way out is to assume that the (quantum) universe is divided
into one part as ``the system to be looked at'' and the remainder as ``the
measuring apparatus'' \cite{FinkelsteinRodriguez1986}. A continuous shift of the
borderline between observing and observed parts of the universe would
then be necessary. In fact, if quantum gravity is to lead to the existence of
a classical limit, i.e., how classical space-time can emerge including 
Einstein's field equation, another part might have to be defined, the 
``environment''. Its wave function is entangled with the measuring part
of the universe (``the apparatus''). The interaction with the environment 
will lead to ``decoherence'' and provide classical properties by a continuous 
measurement process \cite{JoosZeh1985}, \cite{Kiefer1988}. 
Possibly, 
measuring apparatus and environment can be made to coincide in the universe. 
For the interpretation of the wave function of the universe, it may be 
unavoidable to employ some version of Everett's interpretation of quantum 
mechanics; in it the splitting of the wave function by a measurement is 
equivalent to splitting the universe into many copies. In each of these copies 
one of the allowed measurement results occurs
\cite{Everett1957}.\footnote{Cf. also \cite{Mukhanov2007}.} Another 
proposal replaces the ``many worlds'' of Everett by a ``many histories'' 
interpretation in which observers making measurements are within
``decohering'' histories of the same universe \cite{Gell-MannHartle1990}. 

Originative cosmology is taking place in our {\it minds} - as pure mathematics 
does. By it, awareness of what could be {\it potentially} real is produced. 
Passage from the potentially to the {\it actually} real requires a linking 
to an {\it empirical basis}. In the example of Bose condensation, the time 
span between the suggestion of the idea and its experimental validation was 
relatively short: it took about 60 years. The agreement among scientists in 
the case of quantum cosmology may take a very much longer time.

\subsection{Make-believe cosmology: the multiverse}
\label{subsection:multiverse}
The conceptually well founded development of quantum cosmology and quantum 
gravity is very removed from the multiverse scenario to be briefly sketched now.
A multiverse is an ensemble of universes. At best, the elements
(``universes'') of the set are generated from some underlying theory, e.g., 
from the ``string landscape'' (see below). At worst, the ensemble is just
assumed to exist. A multiverse can be represented by a higher-dimensional 
space-time with four or more {\em space} dimensions. Often, this is done within the
framework of ``braneworld'', in which a 3-dimensional space resides in a higher 
dimensional space, called ``the bulk'' to which time is added. Gravitation can
play in the bulk, all other interactions are restricted to the brane. The 
additional spatial dimensions may be compactified or not. The multiverse 
can also consist of an infinite number of replica of one and the same universe
as the many-worlds interpretation of quantum mechanics would imply. Another
case is the multi-domain multiverse with its ``universe-bubbles'' bifurcating 
away from another in particular inflationary schemes (eternal inflation). For a
discussion of different brands of multiverses cf. \cite{Tegmark2004}.

\subsubsection{Multiverse models}
\label{subsubsection:multimod}
The multiverse-concept is introduced in order to help solving philosophical 
problems inherent in, or superimposed on cosmology. With the first, avoidance 
of the singularity at the big bang is meant, with the second an attempt at 
bringing the biosphere back into the realm of the universe (anthropic principles).

In a special approach in brane cosmology, the {\em ekpyrotic model}, the 
universe is embedded as a 3-(mem)brane in a higher-dimensional space plus time
along with other universes (``parallel branes''). All expand independently 
according to general relativity. The ekpyrotic model hypothesizes that the 
origin of the observable universe occurred when two parallel branes collided 
\cite{Khouryal2001}. It is the precurser to cyclic universe models
\cite{Khouryal2004}. In them, a periodical big crunch is followed by a big
bang with up to trillions of years ($\sim 10^{12}$) in between each bang and 
crunch. Density and temperature remain finite. The cyclic universes are said
to be an 
alternative to inflation; they produce the right density fluctuation spectrum 
\cite{Khouryal2002}. A further example for a multiverse scenario is
the so-called ``string landscape''. It is the energy-``manifold''
formed by all degenerated string vacuum solutions (their number is
given as of the order of $\sim 10^{500}$). From each vacuum state a universe is
assumed to ``nucleate'' with a certain probability. Relying on an
estimate ascribed to R. Penrose (\cite{Penrose2005}, p. 728-730), the
nucleation of ``our'' universe (at energies $\sim 10^{16}~GeV$) would
have had only a probability of $10^{-10^{123}}$. 

\subsubsection{Philosophical issues}
\label{subsubsection:philissue}
If all this is not solely forming a mental construct, not just  
philosophers might have difficulties in relating the multiverse with the
notion of ``all that exists in a physical sense''. M. Rees is reducing the 
problem to a semantical one: what we now call ``universe'' could be named 
``metagalaxy''; the 
``multiverse'' would be re-named ``universe'' (\cite{Rees2007}, p. 57). This 
stand hides a change in ontology: the multiverse is taken to exist in the same 
sense as the solar system does. In a correspondence about whether Everett's 
``many-worlds'' interpretation of quantum mechanics should be taken as 
describing infinitely many ``really existing'' universes, or only logical
mental possibilities, B. DeWitt sided with the first claim and asked: ``Is
there any difference'' between things ``physically real'' and ``abstractions 
such as numbers and triangles''? (\cite{Gardner2003}, p. 10). In this spirit, it 
has been claimed recently that the introduction of the concept multiverse is 
leading to ``an extension of the Copernican Principle'': ``The universe is 
not at the center of the world (the multiverse)'' \cite{Mersini-Houghton2008},
p. 13). We cannot but conclude that, in the mind of the author, the multiverse 
now is ``all that exists in a physical sense''. A little less daring was, two 
decades ago, Tipler's definition of the Universe (with a capital U) to consist 
of all logically possible universes where ``Universe'' was the totality of 
everything in existence and ``universe'' a single Everett-branch
\cite{Tipler1986}, \cite{BarrowTipler1986}. Enthusiasm and playfulness may
have seduced some theorists to act on a quip, heard occasionally,: ``All that 
can be thought of and expressed by a mathematical scheme must be realized
in nature, somewhere''. The ``realistic'' view of the multiverse leads
to the uneasy task of finding a link between this system and empirical data 
upon which physics as we know it is based. A task which may well be 
impossible to fulfill (Cf. \cite{Ellis2007b} p. 406). It is not made easier by 
the fact that in many of the multiverse definitions, their universe-elements are 
causally disjoint: they cannot be observed from our place. Apparently,
on the assumption that quantum mechanics is valid also in the multiverse and
that the wavefunctions of the universe-elements can form an entangled
state, we are offered imprints of the multiverse on CMB in the form of
two underdense regions (voids) one of which is connected with the cold spot 
(\cite{Mersini-Houghton2008}, p. 8-9).

A regress ad infinitum is not excluded, with its first step being the 
introduction of the concept ``multi-multiverse'' as the set of all 
multiverses.\footnote{The plural ``multiverses'' has already been amply used, 
albeit only as a logical possibility, not as ``reality''. Cf. several articles 
in \cite{Carr2007} with (\cite{Aguirre2007}, p. 368) as an example.} 

\subsubsection{Multiverse questionaire}
\label{subsubsection:multiquestion}
\noindent The questions asked within the multiverse scenario are quite
different from those of ``quantum cosmology'' (section \ref{subsection:concoquestion}), 
or ``physical cosmology'' (section \ref{subsection:question}). We list some of them:\\
 - How large is the multiverse (finite, infinite)?\\
 - What is its precise structure?\\
 - Do all members have the same (or similar) properties (dimension, geometry,
   physical laws)?\\
 - How can the members be compared (i.e., empirically, not just by a 
   mathematical classification)?\\
 - Is the multiverse (as an ensemble) a dynamical system (with a history), or
   not?\\ 
 - Why is there a need for a selection principle leading to a particular 
   universe?\\
 - How can the values for the (dimensionless) physical constants be derived from
   the multiverse?\\ 
 - Can the multiverse provide the initial conditions for a universe like 
   ``ours''?\\
While, previously, cosmologists were satisfied with trying to find out whether
the fundamental physical constants are depending on cosmic time, or not, now
the demand is to explain why they have the particular values observed
\cite{Tegmarkal2006}. Cosmological modeling is transformed into a bird's eye 
view of the universe: scientists working in multiverse theory seemingly put 
themselves ``outside'' of ``their'' universe (mentally, that is). The necessary
fine-tuning of some of the parameters required for life to exist seems to be 
a strong motivation for the concept of multiverse. It appears to me that many of
the above questions are meaningless within physics; at this time, they
seem to belong into philosophical thinking about the cosmos.

\section{The science of cosmology}
\label{section:cosmoscience}
We have seen that cosmology shows features of descriptive astronomy,
explicatory astrophysics, palaeontology, history, mathematics, physics, and 
natural philosophy. As long as it is handled as {\it cosmophysics}, i.e., as 
an extension of physics from the galactic through the extragalactic realm to 
ever larger massive gravitating structures, it is part and parcel of physics 
proper. Questions relating to parts of the cosmic picture are debated like
those in other branches of physics; an example would be given by the three 
methods for determining baryonic acoustic oscillations \cite{Rassatal2008}. 
The evolution in (past) time is more problematic. As soon as a 
description of the universe (``the world as a whole'') 
by a cosmological model is attempted, knowledge gained is of a ``softer'' 
character than knowledge from astrophysics and planetary science research. 
Synge's statement of the mid 60s, i.e.,  that ``of all branches of modern 
science, cosmological theory is the least disciplined by observation'' 
\cite{Synge1966}), must be shifted nowadays to the inflationary model, quantum 
cosmology and to string theory, though. To what degree can we trust
  in cosmological modeling, to its more than merely descriptive
  imaging of the universe? In view of the necessary correction of the
  distance scale which occured in the 1960s, and of the sudden change
  from $\Lambda=0$ to a non-vanishing contribution of the 
cosmological constant in the 1990s, it should come not as a surprise when 
scientists from other quarters will keep reserved, a little. This applies 
especially to the concept of dark energy. 

\subsection{The epistemic value of cosmology}
\label{subsection:cognitive}
The most characteristic feature of research in the natural sciences is the 
collection of precise empirical data and their connection by self-consistent 
theories. In consequence, technical applications, possible derivation of novel 
relations among the empirical data (``new effects'') obtain as well as models 
of explanation and understanding for the systems investigated. It is essential 
that such explicatory models map, with a minimum of hypotheses, a larger
piece of the network of relationships found in the external world into
percepts of our mind. It is particularly important that we are lead,
by such understanding, to new possibilities of qualitative or, better, 
quantitative experimentation/observation. In view of such demands, is 
cosmological theory represented by the $\Lambda$CDM-model simple, empirically 
well based and conceptually clear? It may be too simple as we will discuss in 
section \ref{subsubsection:errors}. Parts of it, among them the large scale
structure and cosmic background radiation, are empirically extremely well 
supported. Other parts are only very indirectly, e.g., the inflationary
scenario. The part concerned with the era right after the big bang 
(quantum cosmology) has not yet come near an empirical foundation. 
Although the range of their validity is unknown, Einstein's equations, 
their homogeneous and isotropic solutions, the methods
to deviate from them (perturbation theory), and the quest for initial
conditions are conceptually very clear. This cannot be said of the big bang 
concept (origin of space and time?) or, rather, of the whole Planck era which
is neither conceptually nor methodically under control. The concept of
inflation is very clear, in principle, but hazy in its technical details,
e.g., during reheating. An application of cosmology, beneficial for society,
is the development of technology for the improvement of observational tools. 
Another very important one is the emergence of an understanding of the world 
(``Weltbild'') independent of a particular society and its cultural
background; it is owed to the disciplining force of the laws of nature. 

\subsection{The explanatory value of cosmology}
\label{subsection:explanatory}
Nevertheless, one might still worry about the significance of knowledge 
produced by cosmological theory, in particular, about the ``explanatory power''
of the standard model. The concept is used here in the sense of a convincing 
reduction to, or a link with simpler {\em established} facts. Have we now 
understood, beyond a mere {\em description}, why, in the modeled evolution of 
the cosmos, first an extreme {\it global} thinning of matter {\em against} 
gravitational attraction had to occur while, subsequently, massive 
superstructures arose from {\it local} condensations against {\it global} 
expansion? Is it clear why the expansion of the universe after an explosive
phase with deceleration parameter $q=-1$ slowed down to $q\simeq
\frac{1}{2}$ and then steped up again to today's $q=-0.7\pm 0.1$ from
type Ia supernovae? Playing it all back to stochastic perturbations of
a quantized scalar field of unknown origin and uncertain dynamics
compensating gravitational attraction by its negative pressure does
not explain enough. The more so as the initial values have to be put
in by hand as long as no convincing theory for the era before inflation 
is available. 

It is difficult, from the theoretical point of view, to make
transparent the web of assumptions, logical deductions, and empirical
input spun by cosmologists if the explanatory value of the cosmological model is
to be evaluated. Hypotheses of differing weight are intermingled as, for 
example, the classical, {\it relativistic, nonlinear} theory of gravitation, {\it 
nonrelativistic} thermodynamics and kinetic theory for massive particles in
perturbation theory, the relativistic Einstein-Boltzmann equation for the 
{\em fluctuations} of photon and neutrino fields, the {\it linear} theory of 
density fluctuations with non-linear complements, quantum field theory in 
curved space (during inflation), quantization of gravitation, nuclear physics 
(primordial nucleosynthesis) and high energy physics (baryogenesis). 
Approximations are made whenever they are needed for a calculation with the
aim of connecting theory and data.

Special case studies could bring more light. A presentation from which
one might try to get an impression of the explanatory value of
cosmological modeling are 
lecture notes by N. Straumann \cite{Straumann2006}, although not written under 
this aspect. In them, all calculational steps from primordial quantum 
fluctuations until how they show up in the acoustic peaks of oscillating
matter describing the anisotropy of CMB are taken. An 8-parameter description 
for density-, velocity- and metric perturbations is used within two different 
2-fluid-models {\em before} (electrons, baryons, photons plus dark matter) 
and {\em after} recombination (electrons, baryons, dark matter plus photons).
\footnote{In this work, it is assumed that dark energy does not contribute to
  the formation of large scale structures. Other authors wish to include dark 
energy perturbations during the matter dominated era \cite{SaponeKunz2009}.} 

The reliability of the empirical data also has to placed under scrutiny. There are
ambiguities in the interpretation of observations of the large scale
structure (redshift surveys) due to selection effects and the evolution
of objects.\footnote{It is notoriously difficult to get reliable distance
  measurements beyond redshift $z=1$.} There still is a discrepancy between 
the value of the Hubble constant $H_0$ claimed by the $\Lambda$CDM-model (cf. 
section \ref{subsection:concord}) and the much lower value $H_0= 62.3 \pm
1.3~ (\pm 4.0)$ based on the high-accuracy distance indicators of the
astronomers \cite{TammSand2008}. Similar problems arise for the large 
angle scale in CMB, or temperature and noise fluctuations \cite{Lial2009}. 

\subsubsection{Comparison with other natural sciences}
\label{subsubsection:comparison}

A juxtaposition of cosmology with other branches of natural science
with the aim to compare their relative explicative strengths is meaningful only in part.
Of special interest are disciplines with historical aspects like
geology, geophysics and paleontology. There, the evolution of systems is also 
modeled, if only on shorter time scales than the cosmological
ones. One could become inclined to believe that knowlegde about the Earth must be 
easier to obtain and be more secure than knowledge about past eras of the 
universe. Yet, this seems not to be the case. An example is the enigmatic 
solid inner core of the Earth, thought to be formed from small nickel-iron 
crystals. Apparently, it is not homogeneous as one might assume, but shows
large scale structures and anisotropy found through seismic waves
\cite{JephcoatRefson2001}. Explanations are still debated (existence of layers 
etc) but, unlike the anisotropies of CMB, it seems unlikely that those in the 
inner core can be explained by small perturbations to an isotropic Earth
\cite{Anderson2002}. Scenarios about the making of an inner planetary core 
seemingly have not yet converged to an accepted standard one as the
inflationary scenario has in cosmological theory.

Why is it that the physics of the Earth`s innermost core cannot be described as
precisely (in terms of error bars) as the physics of the universe
reflected by the concordance model? A tentative answer would be that the
physics of the universe gets simpler the further we look back into the
past. Simpler than solid state physics applied to the Earth with
its many-body interactions, collective phenomena, phenomenological interactions,
complicated phase transitions. This view is supported by the fact that
the inner core of the {\em gaseous} Sun apparently is known much better. But, is
it exluded that the apparent simplicity of the universe is due to the 
simplifying assumptions underlying the cosmological model and not an
intrinsic feature of the cosmos? In fact, the $\Lambda$CDM-model including
inflation is built in such a way that the imprints of inflation may be seen 
in CMB, but that the microwave background cannot show traces of the ensuing
eras before the last scattering surface. A weaker argument might be that the 
rate of change in the cosmos, after the formation of large structure, is
smaller than in geology. In the inner core of the Earth ``one might expect to 
see changes on a human scale'' \cite{Anderson2002}. 

A similar situation prevails in palaeontology, in which, as in
cosmology, many disciplines like physics, geology, anatomy, technical
mechanics, and biology work together. Here, the evolutionary history
of the Earth including its biosphere is studied. As an example, fossils, say of 
feathered dinosaurs of various periods (in the range of million years
duration), are compared. Phylogenetic trees are constructed with the
help of mathematics. The discovery of an iridium-rich layer at the 
Cretaceous-Tertiary boundary \cite{AlvarezAlvarezal1980} and the ensuing 
suggestion of an asteroid impact as its cause, were tentatively combined to
unravel the mystery of the observed event of mass extinction (of the
dinosaurs), ca. $65\cdot 10^{6}~y$ before the present.\footnote{This
  dating remains virtually unchanged since the 1960s.} Does this idea have an 
assimilable explicatory power as the idea of an inflationary period of the
universe, even if it cannot be expressed within a mathematical model? Aren't 
the ``standard candles'' used in observational cosmology comparable to
fossils? Perhaps, the success with solar nucleosynthesis led us
  to believe that we know more of the physics of supernovae millions of
  light years away than what is known about the touchable fossils of
  palaeontology.

\subsubsection{Error bars}
\label{subsubsection:errors}
The statistical errors of a few percent given by ``precision cosmology'' 
are amazing (Cf. \ref{subsection:concord}). These numbers are reliably
calculated by the best methods available (after filtering and averaging of the 
primary data). Thus, on the one hand, they stand for the progress made in 
assessing the data. In this context, the increased use of methods of Bayesian 
statistics is notable \cite{Trotta2008}. On the other hand, how significant
then is the
uncertainty of $\sim 1\%$ for the age of the universe? It is roughly the same 
uncertainty as presented for the age of the Earth \cite{Dalrymple2001} or, for 
the material from which it was formed \cite{Amelinal2002}. Should't the
absolute dating become more and more precise, the {\em less} we go
back in time? Yet, absolute (chronometric) dating in
palaeo-anthropology tends to be no better than dating in cosmology:
the first appearance of hominids is claimed to be $(7.0~\pm
0.2)~10^{6}~y$ by help of $^{10}Be/~^{9}Be$-dating of the surrounding 
sediments \cite{Lebatardal2008}.  An answer could be that the
  limits in accuracy are set by nuclear physics (radiometric dating),
  i.e., by a precise knowledge of half-lives and decay constants. The
  errors vary from $0.1\%-1\%$ (uranium) to $ \leq 10\%$ (potassium-argon). In
addition, uncertainties from geochemistry (distribution of isotopes) and from 
isotope-chronostratography (changes in the environment needed for the
calibration of radioactivity data) must be added. Dating errors in 
palaeo-anthropology thus cannot be much better than dating in primordial 
or stellar nucleosynthesis. For uncertainties in big bang
nucleosynthesis cf. \ref{subsubsection:nucleo}.

There is a discrepancy between the precision presently ascribed to 
cosmological parameters (errors of 1\% to 10\%) and the lack of 
{\em qualitative} knowledge. 
Quantitatively, the time of (photon) decoupling (via CMB) is set at
$380081^{+5843}_{-5841}~y$ after the big bang (cf. \cite{WMAP2008}, Hinshaw,
G, Weiland, J.L. et al., p. 45, table 7). Can this compensate the fact
that we know less about the much later formation-details of luminous
galaxies {\em near} to us? Although it is widely believed that their 
nuclei house massive black holes, neither by theory nor by simulations, an 
understanding of black hole galaxy seeds has been reached \cite{Madaual2009}. 
The same holds for spiral galaxies with thin disks. The 
$\Lambda$CDM-model can give only a relatively crude picture of structure
formation and evolution. But perhaps, this is the domain of astrophysics, not
of cosmology. Simulations of galaxy formation and evolution have met with
great success, cf. \cite{SpringelWhiteal2005}. Similarly, the age at
reionization is given to be $432^{+90}_{-67}~\times 10^6~y$. The hope is that
plasma physics at that time has been understood well enough and that its 
consequences for CMB have been taken into account (cf. \cite{Opher1997}, 
\cite{Mukhanov2005}, p. 407-409). For the {\em cognitive} value of a physical 
model numerical precision does not play the decisive role. However, numerical 
precision has to be taken dead serious for predictions into the future. The 
precise numbers produced by CMB within the $\Lambda$CDM-model are very
relevant if alterations of the cosmological model will be
attempted. However, they are as irrelevant to society with regard to
the future as are the ages related to palaeontology. Progress of
precision cosmology reflected by the narrowing of error bars may be of
an {\em intra-theoretical} value, only.

\section{Conclusion}
\label{section:conclusion}
Throughout history mankind has tried to picture the world and to
understand its origin and its features (Cf. \cite{Kragh2007}). In
  cosmology as a very young scientific discipline, ideas of different 
scientific quality are encountered. In this situation, opinions seem
to play a prominenter role than in some other parts of
physics. Nevertheless, today, through the $\Lambda$CDM-model, physical 
cosmology provides an image of the universe not in conflict with the
wealth of data gained by painstaking observation and intelligent
theoretical interpretation. The achieved scientific description of
``the world as a whole'' is a remarkable asset independent of a
particular cultural background. Nevertheless, the question asked in
the title can receive only a guarded answer: As described in this
paper, in view of the haziness of the universe's 
extension in time and space, and because of the methodological and epistemic 
problems of cosmological modeling, knowledge gained about the ``world
as a whole'' cannot be as secure and explicative as knowledge from 
laboratory or planetary physics. Silk called cosmology a {\it
  falsifiable myth} \cite{Silk1987}. Certainly, a tremendous amount of 
additional empirical data concerning the large scale structure obtained since 
has been used to strengthen the cosmological model. Yet, with almost all of
the universe's matter content unexplained, the situation still is the same: We 
modestly conclude that mathematical modeling, in particular when dealing with 
the early and earliest epochs of the universe, cannot produce but the 
cosmological myths adequate for our time.

\section{Acknowledments}
For his very helpful and detailed comments, and for suggesting improvements to
some incorrect statements in a previous version, I am very grateful to
D. J. Schwarz, Bielefeld. I also thank my G\"ottingen colleagues from
astrophysics, F. Hessman, and from geophysics, A. Tilgner, for
discussions and for some references.  



 \end{document}